\documentclass[twocolumn,pre,aps,superbib,tightenlines,floatfix,superscriptaddress,showpacs]{revtex4}
\usepackage[T1]{fontenc}
\usepackage{ae,aecompl}

\usepackage[latin9]{inputenc}
\usepackage{amsmath}
\usepackage{graphicx}
\usepackage{amssymb}
\usepackage{hyperref}
\usepackage{cases}
\usepackage{enumitem}

\begin{document}

\title{Self-organization and solution of shortest-path optimization problems with memristive networks}

\author{Yuriy V. Pershin}
\email{pershin@physics.sc.edu} \affiliation{Department of Physics and Astronomy, University of South Carolina, Columbia, South Carolina 29208, USA}
\author{Massimiliano Di Ventra}
\email{diventra@physics.ucsd.edu} \affiliation{Department of Physics, University of California, San Diego, California 92093-0319, USA}

\begin{abstract}
We show that memristive networks--namely networks of resistors with memory--can efficiently solve shortest-path
optimization problems. Indeed, the presence of memory (time non-locality) promotes self organization of the network
into the shortest possible path(s). We introduce a network entropy function to characterize the self-organized evolution, show the solution of
the shortest-path problem and demonstrate the healing property of the solution path. Finally, we provide an algorithm to solve the traveling salesman problem. Similar considerations apply to
networks of memcapacitors and meminductors, and networks with memory in various dimensions.
\end{abstract}

\maketitle

\section{Introduction}

Currently, much attention is being devoted to memory circuit elements (memelements) \cite{diventra09a,diventra09b} and their applications in different domains of electronics \cite{pershin11a}.
In particular, computing with memelements \cite{pershin12a,diventra13a} -- {\it memcomputing} -- has received much attention because of the ability of network architectures built out of memelements to store
and process information on the same physical platform \cite{pershin12a,diventra13a,borghetti10a}.
This is a major conceptual and practical departure from the present day architectures based on the von Neumann machine concept \cite{von1993first}. The strength of memcomputing is essentially based on the {\it analog massively-parallel dynamics} of many--if not all--memelements in the network and the ability to recover the result(s) of the computation from the same computing units, much like the brain is thought to operate \cite{pershin11d,diventra13a}. Examples of approaches to compute with memelements include
neuromorphic computing with memristive synapses \cite{snider08a,Barranco09b,Afifi09a,pershin10c,jo10a,Ebong12a,pershin12a}, massively-parallel computing with memristive networks \cite{pershin11d,ye2013computing}, logic with memory circuit elements \cite{strukov05a,Lehtonen09a,borghetti10a,pershin12a}, and memristive cellular automata \cite{Itoh10a}.

In order to realize this memcomputing paradigm some criteria need to be satisfied. We have introduced and expended on these criteria in Ref.~\cite{diventra13a}. These are
\begin{enumerate}
\item Scalable massively-parallel architecture with combined information processing and storage.

\item Sufficiently long information storage times.

\item The ability to initialize memory states.

\item Mechanism(s) of collective dynamics, strong "memory content".

\item The ability to read the final result (from relevant memelements).

\item Robustness against small imperfections and noise.
\end{enumerate}

One can show that the memcomputing schemes mentioned above satisfy all or the majority of these
criteria.~\cite{diventra13a} We also note that although memristive devices \cite{chua76a} are the main focus of present efforts, memcapacitive and meminductive devices \cite{diventra09a} are also of great promise because of their potential to perform calculations at a minimal energy cost.

Previously, we demonstrated a maze problem solution using memristive networks \cite{pershin11d}. Clearly, all of the above requirements are satisfied within this approach \cite{diventra13a}.
In this paper we consider a more complex case of the shortest-path optimization -- the shortest path optimization on a two-dimensional plane. In fact, our approach and results can be easily extended to
any hyperplane in an $N$-dimensional space, with $N>1$. The difficulty inherent in this problem stems from the fact that any two points of the plane are connected by multiple often degenerate paths, which complicates the finding of single optimal solution. We show that random networks --without any predefined symmetry directions-- are more promising in this regard. In addition, we suggest an algorithm that provides a possible solution to the NP-hard traveling salesman problem. A further development of the suggested algorithm may lead to some quite interesting results on the long way \cite{applegate2011traveling} toward finding more efficient solutions of this problem.

The paper is organized as follows. In Sec. \ref{sec1}, we discuss essential details of memcomputing schemes utilizing memristive networks. Sec. \ref{sec2} presents our studies of memcomputing with symmetric (regular) networks. In particular, we show that regular networks can find a unique solution if a pair of specified nodes is located along a symmetry direction. In a more general case, however, a doubly degenerate solution is obtained. Moreover, we demonstrate the healing property of the solution path (Sec. \ref{sec22}).  Sec. \ref{sec3} studies memcomputing with random networks suitable to solve the shortest-path problem in any direction (Sec. \ref{sec31}). We also suggest an algorithm that provides an approximate solution to the traveling salesman problem (Sec. \ref{sec32}). Finally, in Sec. \ref{sec4} we give our conclusions.

\begin{figure}[t]
 \begin{center}
    \includegraphics[width=5cm]{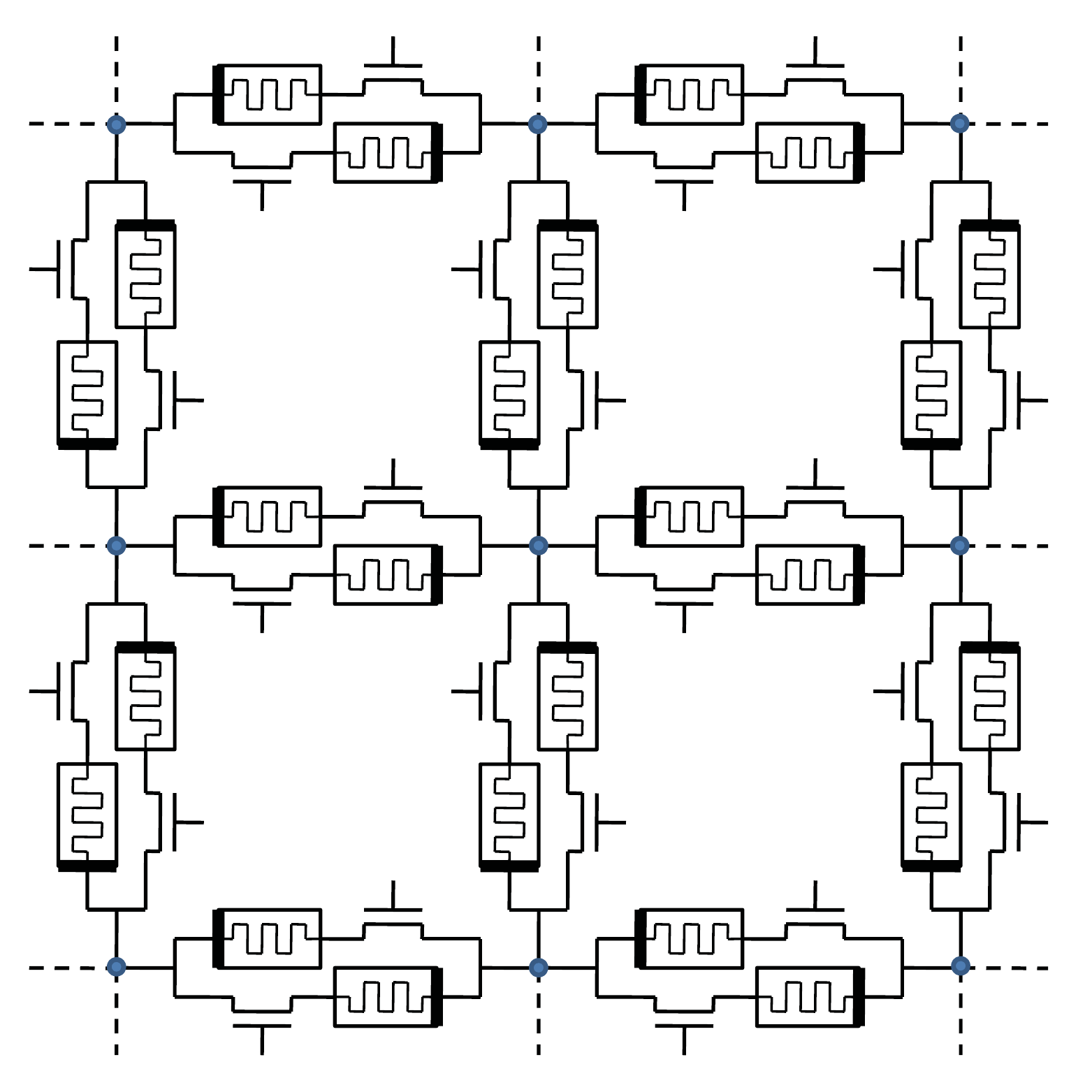}
\caption{\label{fig1} An example of memristive processor consisting of a network of memristive elements in which
each grid point is attached to several basic units. Each basic unit involves
two memristive devices connected symmetrically (in-parallel) and
two switches (field-effect transistors). The switches provide access to individual memristive devices, while
in-parallel connection symmetrizes the response of bipolar memristive elements.
}
\end{center}
\end{figure}

\section{Memcomputing with memristive networks} \label{sec1}

\begin{figure*}[t]
 \begin{center}
  \centerline{
    \quad \quad \quad \quad \quad
    \mbox{\includegraphics[width=7.00cm]{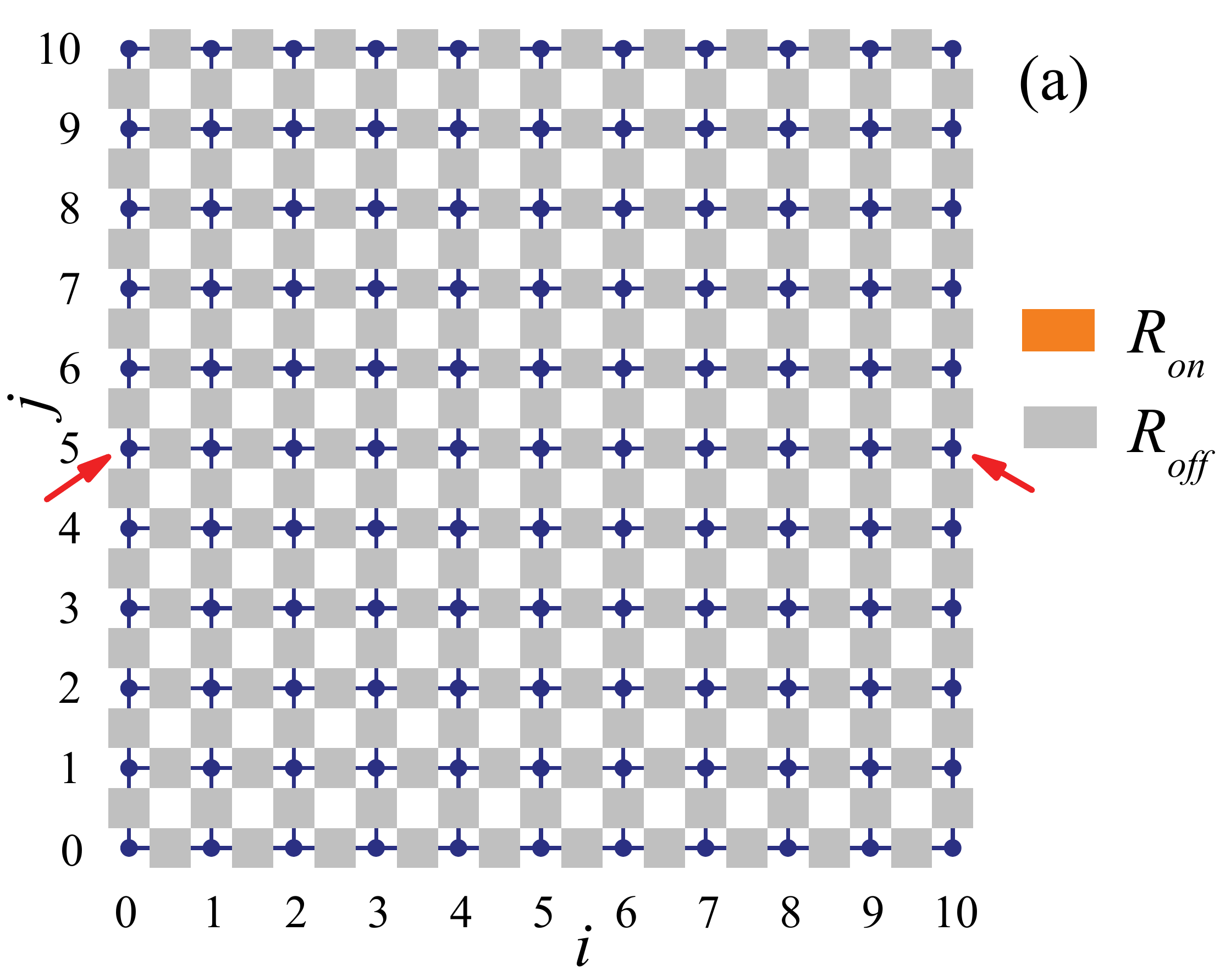}}
    \quad \quad
    \mbox{\includegraphics[width=7.00cm]{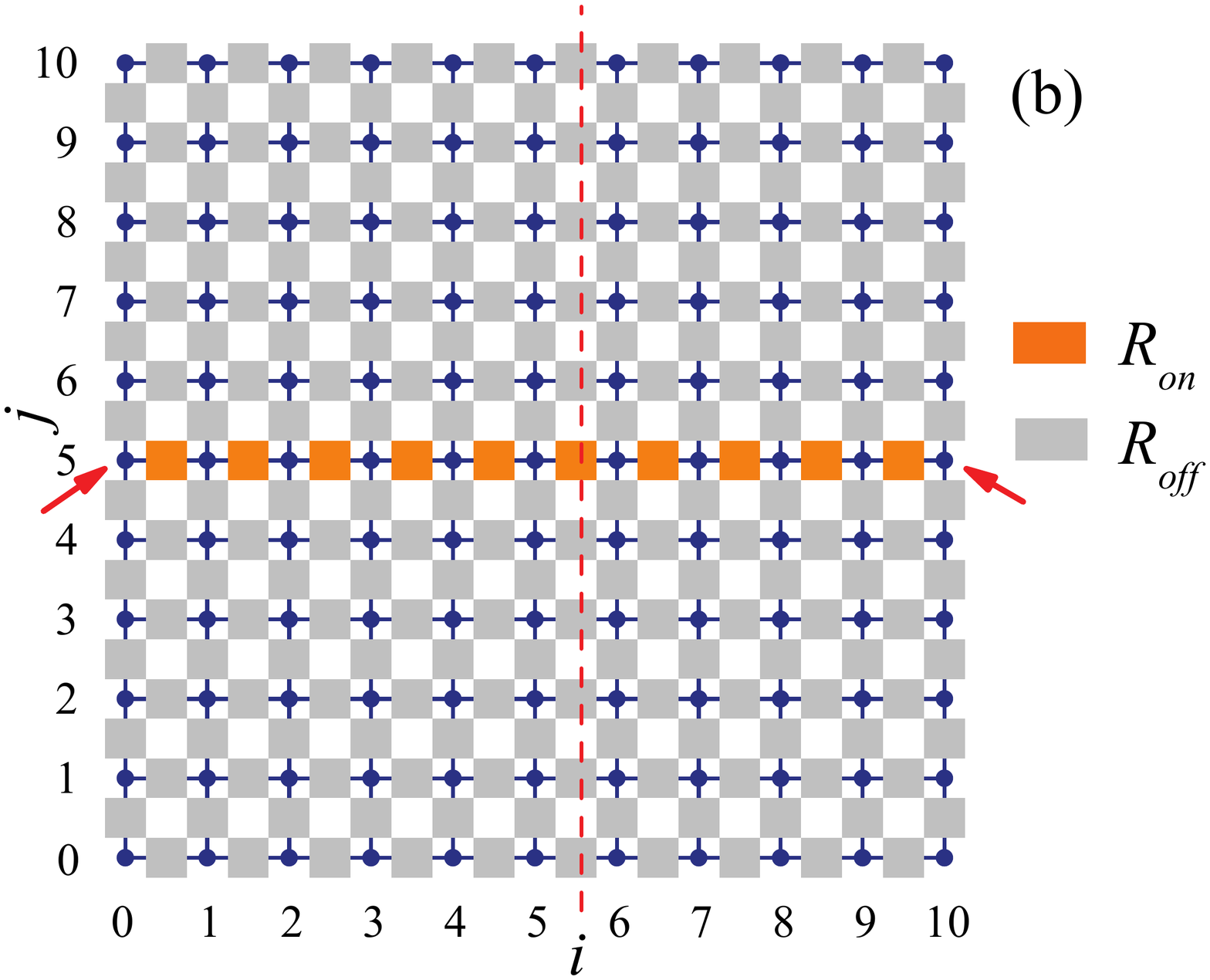}}
  }
  \centerline{
    \mbox{\includegraphics[width=7.00cm]{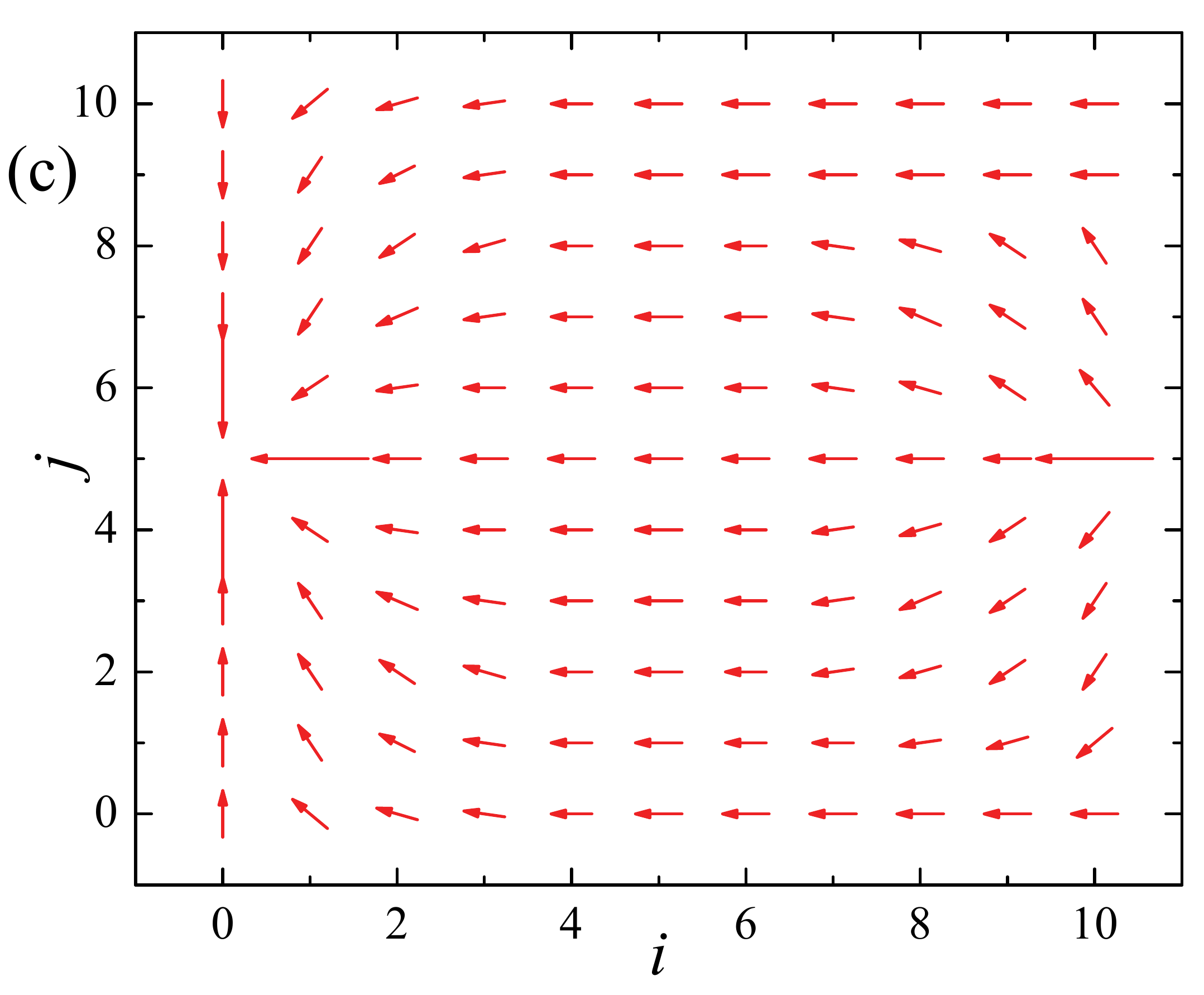}}
    \quad \quad
    \mbox{\includegraphics[width=7.00cm]{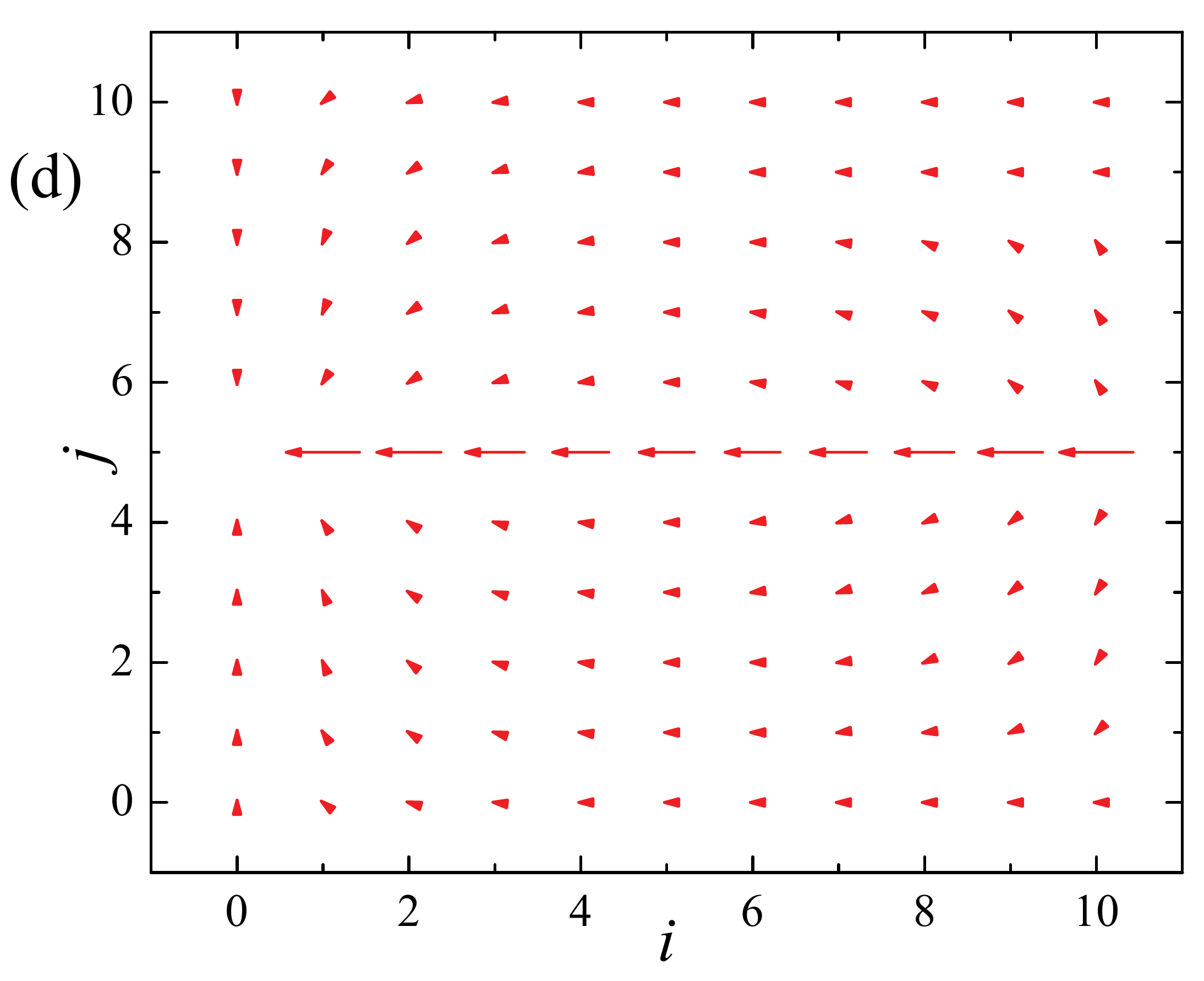}}
  }
 \caption{\label{fig2} (Color online). Solution of the shortest path problem for the pair of nodes indicated by red arrows in (a) and (b) in a an 11$\times$11 memristive network. (a) Initial and (b) final states of the memristive network. Here, the memristance of each basic unit (involving two memristive devices) is represented by a color. The vertical red (dashed) line shows the network cross-section used in entropy calculations. Distributions of electron current corresponding to (a) and (b) are shown in (c) and (d), respectively. See
 Methods for details of all calculations. }
\end{center}
\end{figure*}

\begin{figure}[t]
\begin{center}
    \includegraphics[width=7.0cm]{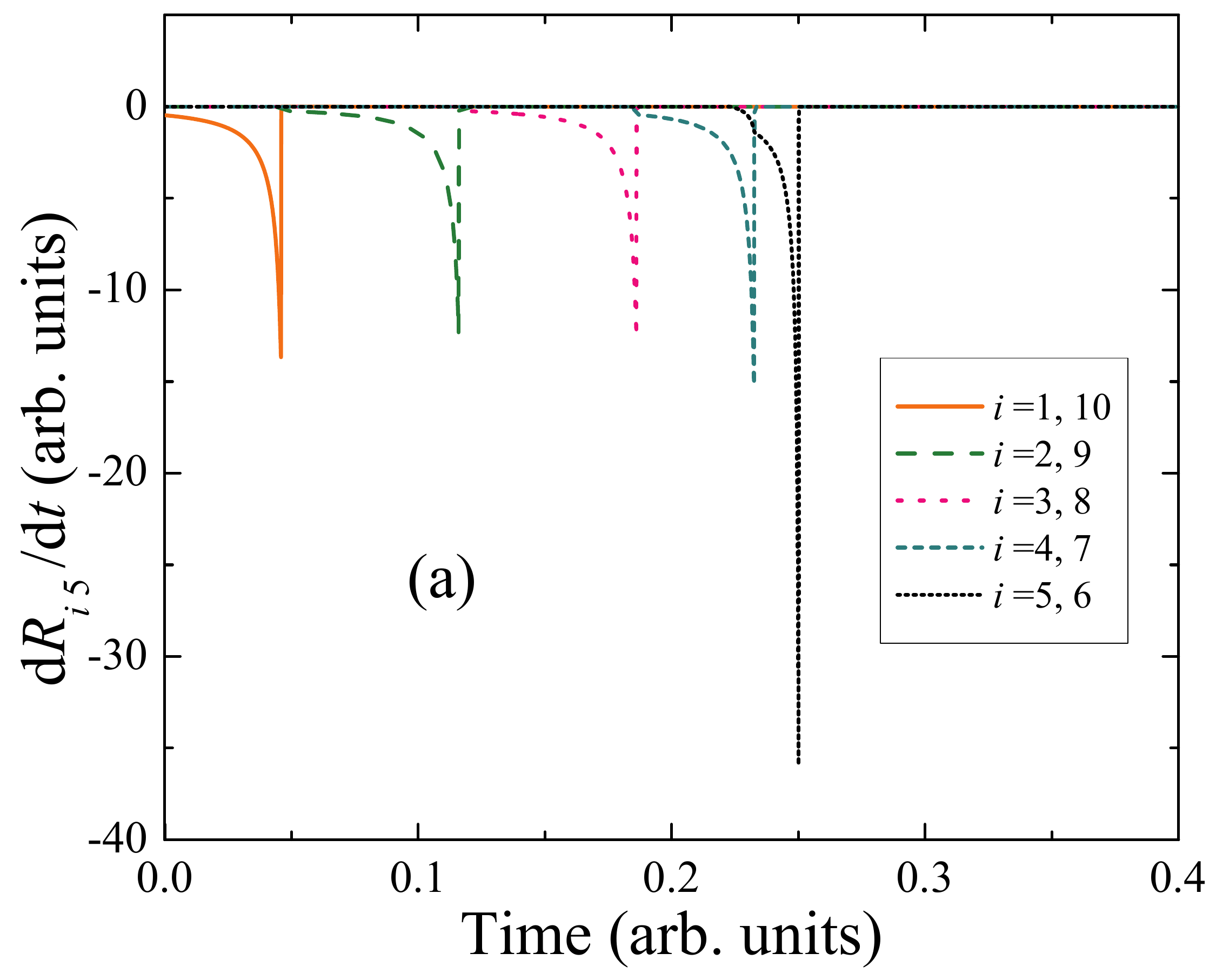}
    \includegraphics[width=7.0cm]{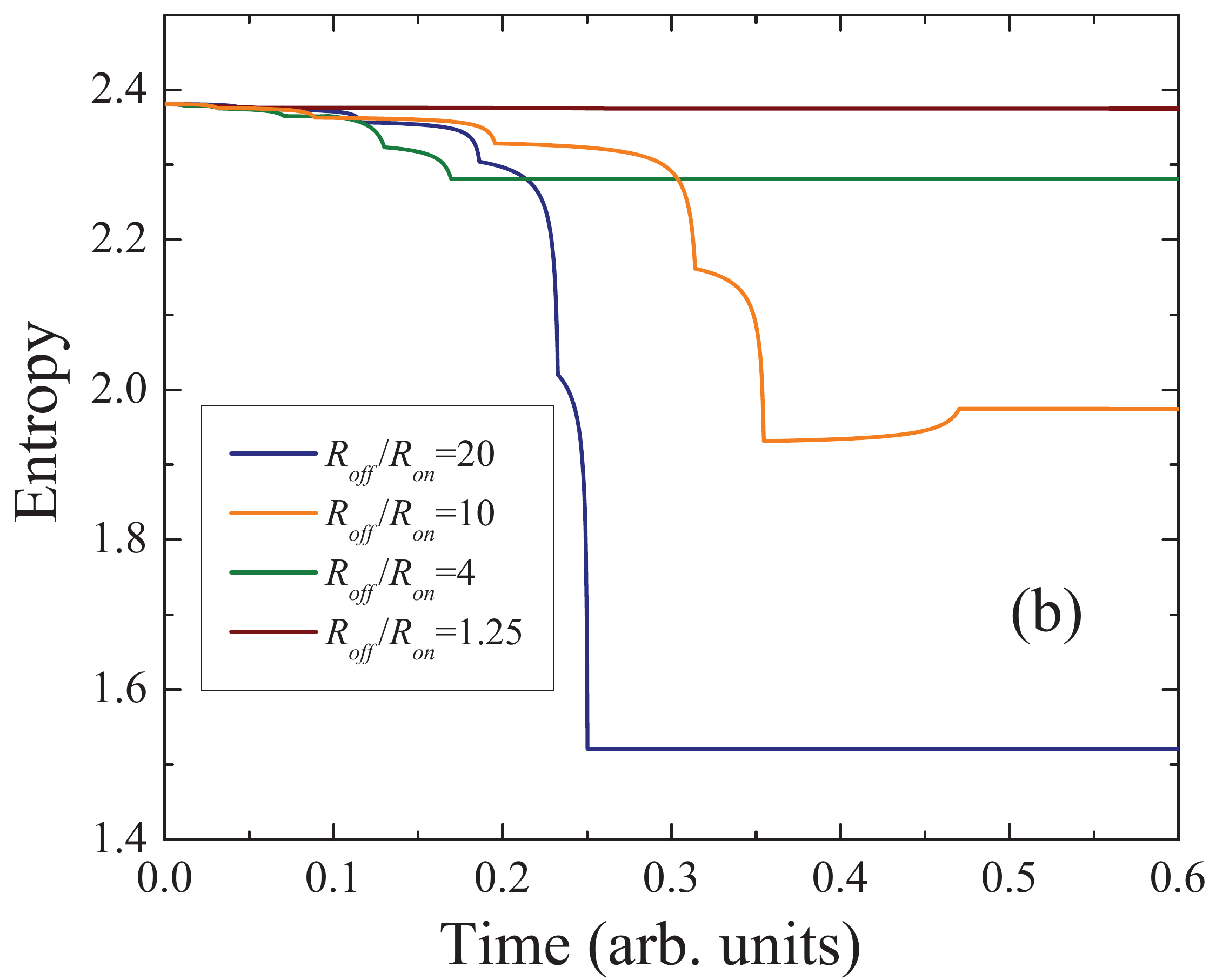}
\caption{\label{fig3}  (a) Dynamics of resistance switching within the calculation stage corresponding to the shortest-path problem solution
presented in Fig. \ref{fig2}. This plot shows that the solution emerges from both sides and propagates to the center.
(b) Network entropy as a function of time from Eq. (\ref{eq_entropy}) for networks of different memory content. A slight increase of the entropy for the $R^M_{off}/R^M_{on}=10$ curve at the time of about 0.4 (arb. units) is due to a delayed switching of four vertical units directly connected to the input/output nodes.}
\end{center}
\end{figure}

\begin{figure}[t]
 \begin{center}
    \includegraphics[width=7.5cm]{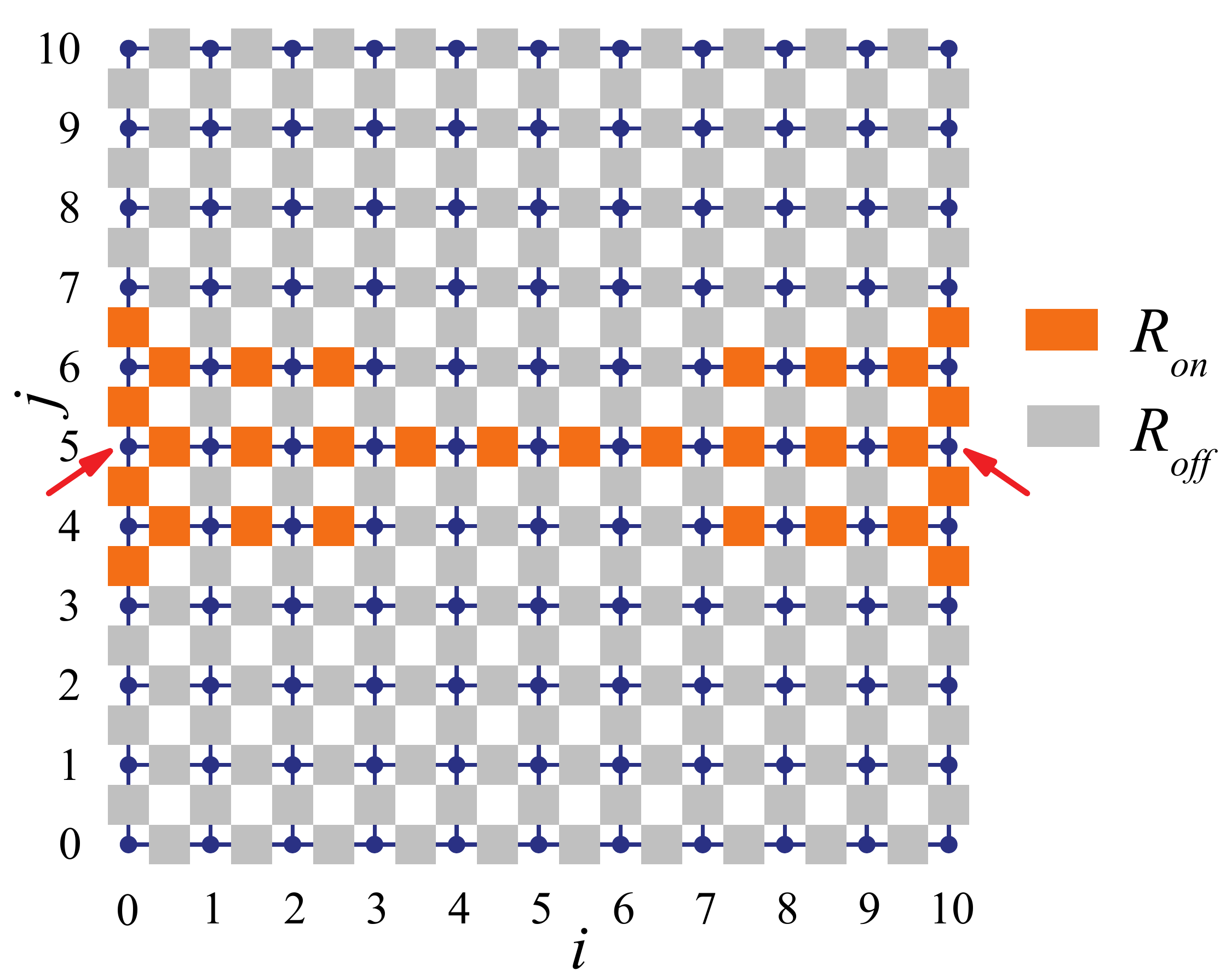}
\caption{\label{fig4} (Color online). Solution of the shortest-path problem by a network of low memory content, $R^M_{off}/R^M_{on}=1.25$.}
\end{center}
\end{figure}

\begin{figure}[t]
 \begin{center}
    \includegraphics[width=7.5cm]{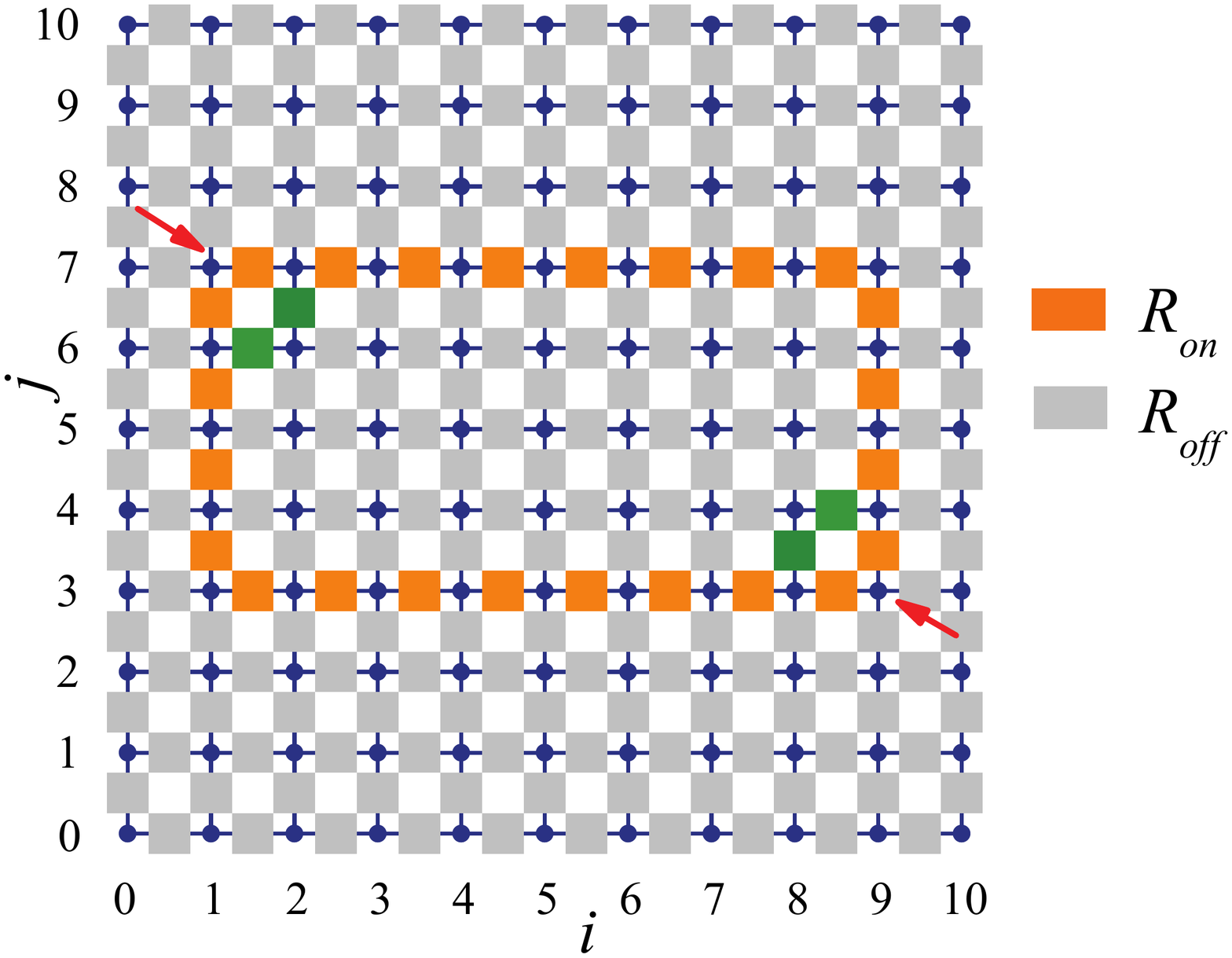}
\caption{\label{fig5} (Color online). Solution of the shortest-path problem by a network of high memory content, $R^M_{off}/R^M_{on}=10$, when the input and output nodes (shown by arrows) are selected not along a symmetry direction. Two possible solutions are found.}
\end{center}
\end{figure}
\subsection{Memristive Processor}

A memristive processor (network) consists of a collection of grid points on an hyperplane of $N$-dimensional space ($N> 1$) connected by
basic units involving memristive elements and traditional elements, such as switches (see Fig. \ref{fig1}). Below, we consider different
organizations of the grid points including a regular square array such as that shown in Fig. \ref{fig1} and a random one on a 2D plane.
Although not strictly necessary, in the examples we have chosen the design of
each basic unit to be symmetric (involving two bipolar memristive
elements) so as to conveniently provide independency of the circuit operation on the sign of the applied voltage. We
also introduce switches for two main functions: {\it i}) to provide independent access to each individual memristive device, and {\it ii}) to define the network topology, if needed (see Ref. \onlinecite{pershin11d} for an example). Thus, the architecture of Fig. \ref{fig1} provides access to each individual memristive device for the purpose of initialization and reading of the calculation result. The calculation consists in the evolution of the network state--defined as the collection of states of all memristive devices--when an appropriate pulse sequence is applied to a sub-set of grid points.

\subsection{Algorithms}

Computation algorithms for memristive processors involve three standard stages: initialization, calculation and result reading. In this work, we focus on solving the shortest path and traveling salesman problems. The corresponding algorithms are formulated below.

\subsubsection{The shortest-path problem} \label{sec121}

In the shortest-path problem, we want to find the shortest path between two specified (input and output) nodes (such as, for example, those shown by red arrows in Fig. \ref{fig2} (a) or (b)) in the network. The algorithm used for this calculation is the following:

{\it Initialization stage}: all memristive devices are pre-initialized into the "OFF" (high-resistance) state.

{\it Calculation stage}: a voltage pulse of a suitable amplitude and duration is applied to the pair of specified nodes.

{\it Calculation reading}: the shortest path is given by a sub-set of basic units in their "ON" (low-resistance) state \footnote{The memristance of a basic unit is that of two memristive devices connected in-parallel. Note that in-parallel connection of two memristive devices can be described as a single higher-order memristive device.}.

\subsubsection{Traveling salesman problem} \label{sec122}

In the traveling salesman problem \cite{applegate06a}, a salesman is given a set of cities to visit. The objective of the salesman is to find a path which minimizes the round-trip distance --each of the cities should be visited only once. In order to solve this problem, we will consider a two-dimensional memristive network and select a set of nodes representing all the cities in the set. The actual geographic locations of the cities will be taken into account for the selection of nodes. The basic algorithm for the traveling salesman problem includes the following steps:

{\it Initialization stage}: all memristive devices are pre-initialized into the "OFF" (high-resistance) state.

{\it Calculation stage}: a sequence of square voltage pulses of random polarity is applied to randomly selected pairs of nodes representing the cities. During this stage, low-resistance paths in the network will start to develop. Importantly, the memristive network will retain memory on the previously applied pulses, and the subsequent network evolution will be largely based on the topology of the already developed low-memristance paths. In this way, an optimized problem solution emerges. Moreover, we anticipate that the path formation in the network occurs {\it hierarchically}: at shorter times, the low memristance paths between closely spaced cities are formed, and, at longer times, these different group of cities become connected. As the electric current tends to flow through the shortest/least resistive path, we expect that the proposed algorithm works out the traveling salesman problem, at least in some cases.

{\it Calculation reading}: the shortest path is given by a sub-set of basic units in their "ON" (low-resistance) state.

\section{Self-organization and healing in regular networks} \label{sec2}

\subsection{Self-organization} \label{sec21}

We now provide an explicit example of computing with memelements -- specifically with regular memristive networks -- in order to exemplify even further the criteria given above, and the possibilities offered by this paradigm. Details
of the specific memristive systems used and the simulation details can be found in Appendix A.

Fig. \ref{fig2}(a) shows the initial state of the network when all memristive devices are in their "OFF" states. At the initial moment of time $t=0$, a single constant amplitude voltage pulse is applied to the input/output nodes shown by red arrows in Fig. \ref{fig2}(a). The final state of the memristive network (the calculation result) is presented in Fig. \ref{fig2}(b). Clearly, two specified nodes are connected by a chain of memristive devices in the "ON" state giving the shortest path problem solution. Note that Fig. \ref{fig2}(a), (b) depict the memristance of each basic unit consisting of two memristive devices.

Let us now consider the network evolution -- the dynamics of the calculation stage -- in more details. First of all, we would like to mention a similarity between the process of computing as performed by the memristive processor and the ant-colony optimization algorithm \cite{Colorni91a,book_ants}. The latter is an adaptable algorithm inspired by the observation that ants, upon finding food, mark their trails by pheromones thus attracting other ants in order to reinforce the trail closest to their nest. A similar type of reinforcement is observed in the memristive network dynamics. Fig. \ref{fig2}(c) shows the current distribution in the network at the initial moment of time when all the memristive devices are in the "OFF" state. In this case, the current flows in multiple paths. However, since the rate of memristance change is proportional to the current, the memristance of the least resistive path will decrease faster ``attracting'' more and more current. Therefore, the current flowing through the least resistive path will reinforce this path, similarly to the trail reinforcement of the ant colony, see Fig. \ref{fig2}(d).

Moreover, it is interesting to note that the problem solution develops gradually, starting from {\it both} the specified nodes. This is clearly seen in Fig. \ref{fig3}(a) that presents the rate of change of the memristances along the solution path as a function of time. Note also that the
reinforcement of the solution is not supervised, thus implying that the shortest path arises spontaneously as a
{\it self-organized process}. This is an important fundamental result: it is the presence of memory--namely time non-locality--in the system that leads to self-organization of the dynamics: no memory, no self-organization.

\subsection{Network entropy}

In order to quantify the system evolution even further, we define a {\it network entropy} with respect to currents in the network (similarly, we can use memristances for this purpose). For example, by considering currents through a vertical cross section of the network at its center (see Fig. \ref{fig2}(b)), we can define the network cross section entropy as \footnote{Note that in the absence of capacitive components the total current is independent of the choice of this cross section at any given time,
provided such cross section crosses the shortest-path solution and spans the far upper and lower ends of the network without self-intersecting.}
\begin{equation}
\sigma_i(t)=-\sum\limits_{j=0}^{N-1}\tilde I_{ij}(t)\textnormal{ln}\left( \tilde I_{ij}(t) \right),
\label{eq_entropy}
\end{equation}
where $N$ is the number of basic units connected in the horizontal direction (crossed by the red (dashed) line in Fig. \ref{fig2}(b)), $j$ is the index of the row of horizontal basic units crossed by the cross-section, $\tilde I_{ij}=I_{ij}/I_{tot}$ is the normalized current through a horizontally connected basic unit, and $I_{tot}=\sum_{j=0}^{N-1} I_{ij}$. Fig. \ref{fig3}(b) demonstrates (with
$N=11$ and $i=6$) that the network entropy decreases in time as the computation proceeds. In statistical physics, the entropy is related to the number of states available to the system. Here, its decrease can be thus interpreted as due to the decrease in the number of paths available for the current, with a more pronounced decrease the larger the memory content in the system (as represented by the ratio $R_{off}/R_{on}$). Alternatively, one can consider an entropy defined for the complete network, namely, $\sigma(t)=-\sum_{ij}\tilde I_{ij}(t)\textnormal{ln}\left( \tilde I_{ij}(t) \right)$, where the summation is performed over all edges in the network. Qualitatively, its time dependence is expected to be close to that shown in Fig. \ref{fig3}(b) for the cross section entropy.

In order to show that networks with higher memory content provides a better solution, we perform a calculation similar to that presented in Fig. \ref{fig2} assuming, however, a much smaller difference between $R^M_{on}$ and $R^M_{off}$ of each memristive device ($R^M_{off}/R^M_{on}=1.25$ compared to $R^M_{off}/R^M_{on}=20$ used in the previous calculation). Fig. \ref{fig4} demonstrates that now the solution of the shortest-path problem can not be found exactly at the
given bias. In fact, in addition to the switching of memristive devices directly connecting the input and output nodes, many other memristive devices are also switched into the "ON" state (see Fig. \ref{fig4}). This example demonstrates the importance of the strong "memory content" requirement (criterion {\bf 4} discussed in the Introduction, see also Ref. \onlinecite{diventra13a}).

In the general case, however, when the selected nodes are located along an arbitrary direction (different from the network symmetry directions), regular networks fail to uniquely find the problem solution. We demonstrate this feature in Fig. \ref{fig5} showing a shortest-path problem solution, which is doubly degenerate and developed along the symmetry directions of the network. This is a direct consequence of the network symmetry. Consequently, we expect that networks without any local or global symmetry--random networks--or ordered networks but not periodic--such as  quasi-periodic networks \cite{levine1984quasicrystals}--are best candidates to solve shortest-path optimization problems on the hyperplane. In Sec. \ref{sec3} below we indeed show that random networks can find the proper solution for arbitrary oriented selected nodes.

\subsection{Healing of the damaged solution} \label{sec22}

\begin{figure}[bt]
 \begin{center}
    \includegraphics[width=7.5cm]{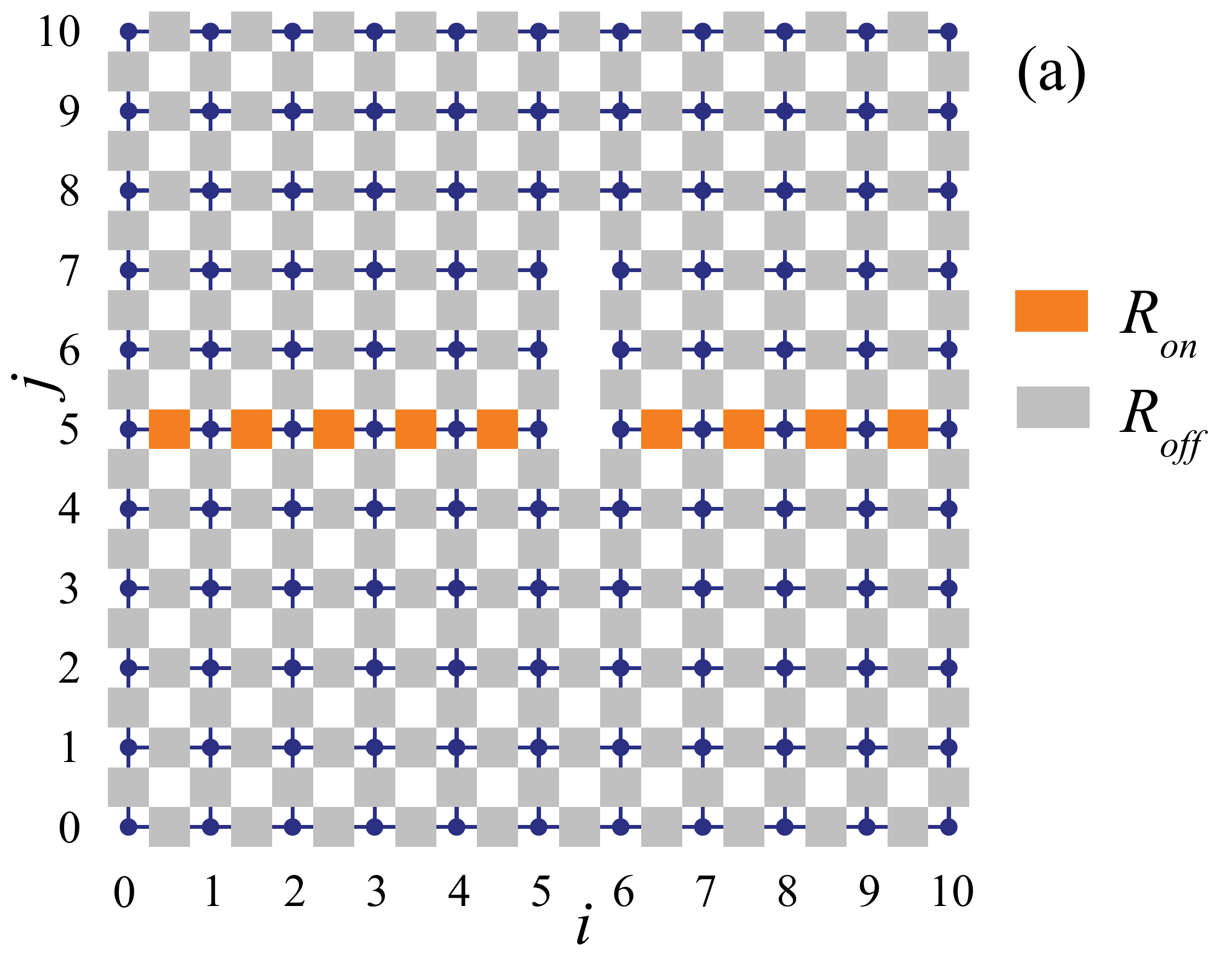}
    \includegraphics[width=7.5cm]{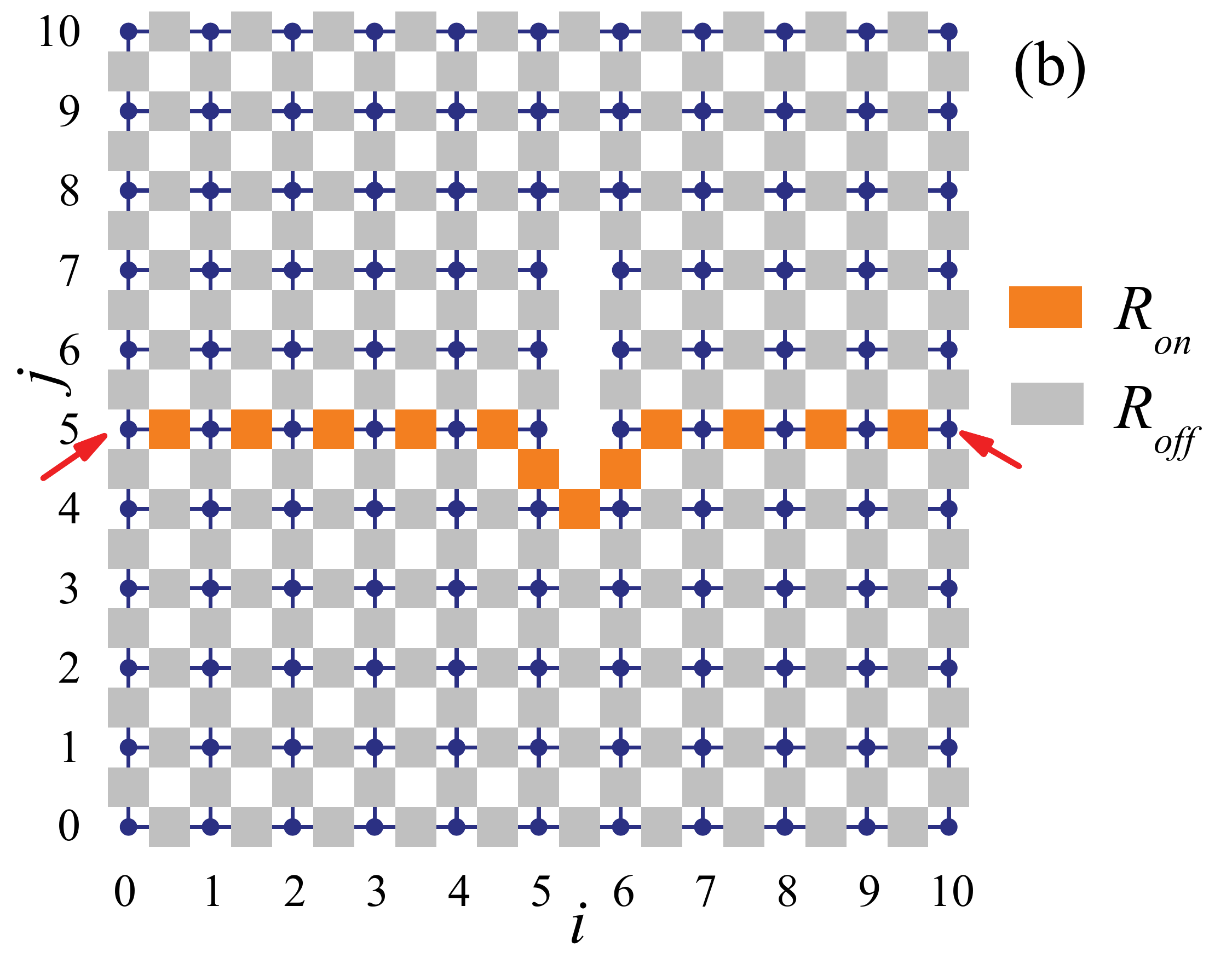}
\caption{\label{fig6} (Color online). Healing (b) of a damaged (a) solution. To heal the solution damage in (a),
a single square pulse of appropriate width and duration is applied to the input and output nodes shown by the red arrows in (b).}
\end{center}
\end{figure}

In order to analyze criterion {\bf 6} discussed in the Introduction in more detail, let us now damage the shortest-path problem solution shown in Fig. \ref{fig2}(b) (as well as the network) by removing three horizontal basic units in the central part of the network as shown in Fig. \ref{fig6}(a). We note that, due to memory, the memristive network has a remarkable ability to {\it repair} damaged solutions--the healing ability we have mentioned above. Indeed, this property is close to the {\it self-healing} ability that can be ascribed to systems or processes, which by nature or by design tend to correct any disturbances.

The healing of the damaged solution is performed by applying a single pulse of a certain amplitude and duration to the input and output nodes. The result presented in Fig. \ref{fig6}(b) shows that a new path connecting two pieces of the initial shortest-path solution develops below the damaged region. The three missing connections have been removed intentionally in an asymmetric fashion in order to show that the healing occurs along the shortest possible path around the damaged region.

It is easy to understand the origin of this healing process: as soon as we switch on the pulse between the input and
the output nodes, the current will flow through all possible paths. However, the shortest one is again the one that is mostly affected, and thus reinforced during dynamics.

The stability of the shortest-path problem solution to small imperfections of the system (e.g., finite width distributions of threshold current, limiting values of memristance, etc.) is evident, and therefore does not deserve a closer inspection.

\section{Random networks} \label{sec3}

Let us now discuss the use of networks with no local/global symmetry in the solution of optimization problems.
Such random networks are generated using the following algorithm. The nodes are placed randomly in a $Na\times Na$ square area where $N$ is a number and $a$ is a unit length  with the constraint that no two nodes can be closer than $0.9 a$. After that, any two nodes located closer than $1.5a$ are connected by a memristive edge. We note that the random networks obtained in this way may involve some local intersections of edges that do not play any significant role in the network dynamics, but in principle could be realized on a chip. Therefore, we do not bother with resolving the crossings. Moreover,
we assume that each edge is of a similar structure to that shown in Fig. \ref{fig1} and, on the hardware level, all operations such as initialization, dynamics, and reading of the final results are performed as described in Sec. \ref{sec1}. Therefore, the only difference between regular and random networks is in the network topology.

\subsection{The shortest-path problem}  \label{sec31}

\begin{figure}[bt]
 \begin{center}
    \includegraphics[width=7.5cm]{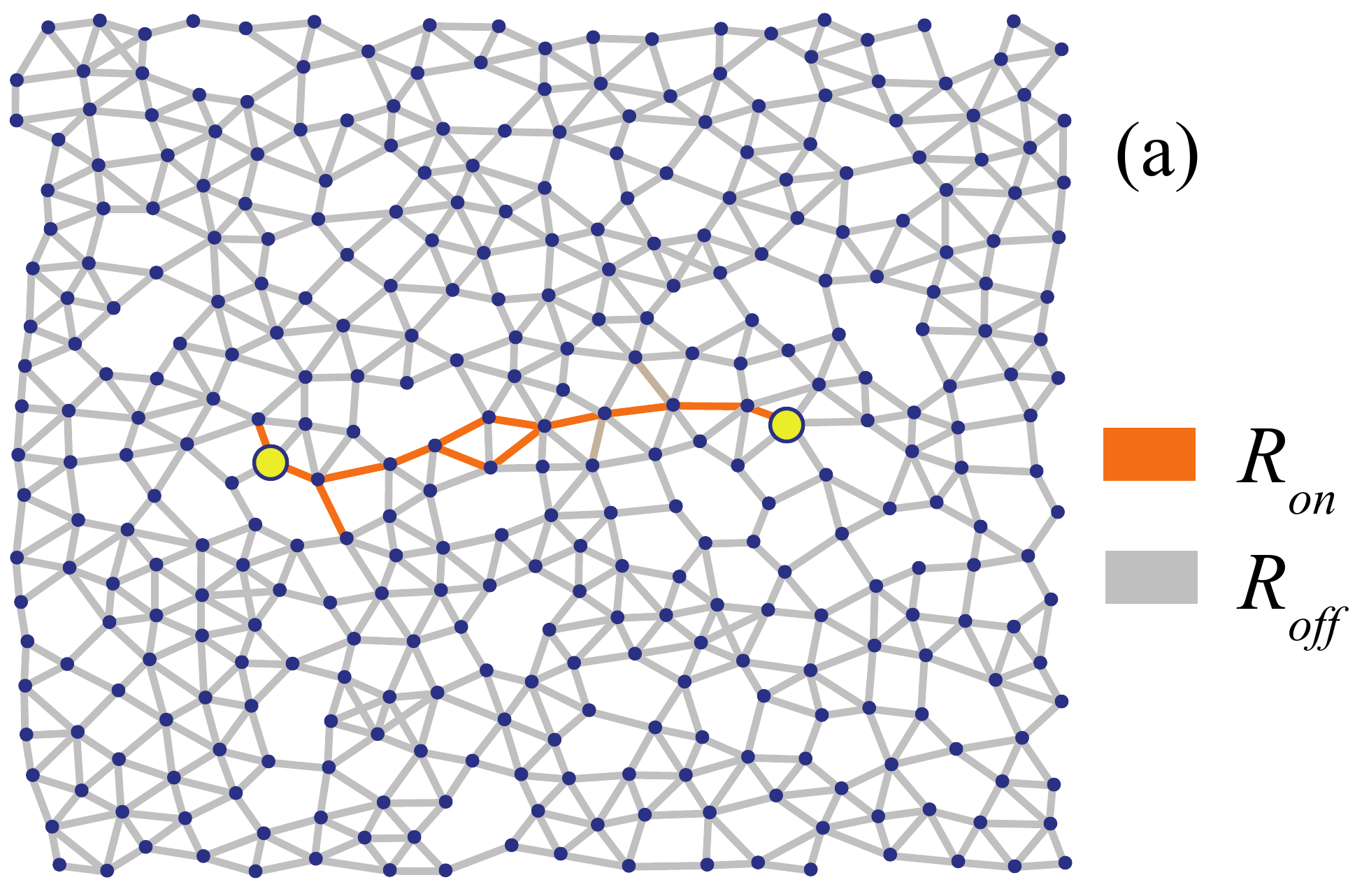}
    \includegraphics[width=7.5cm]{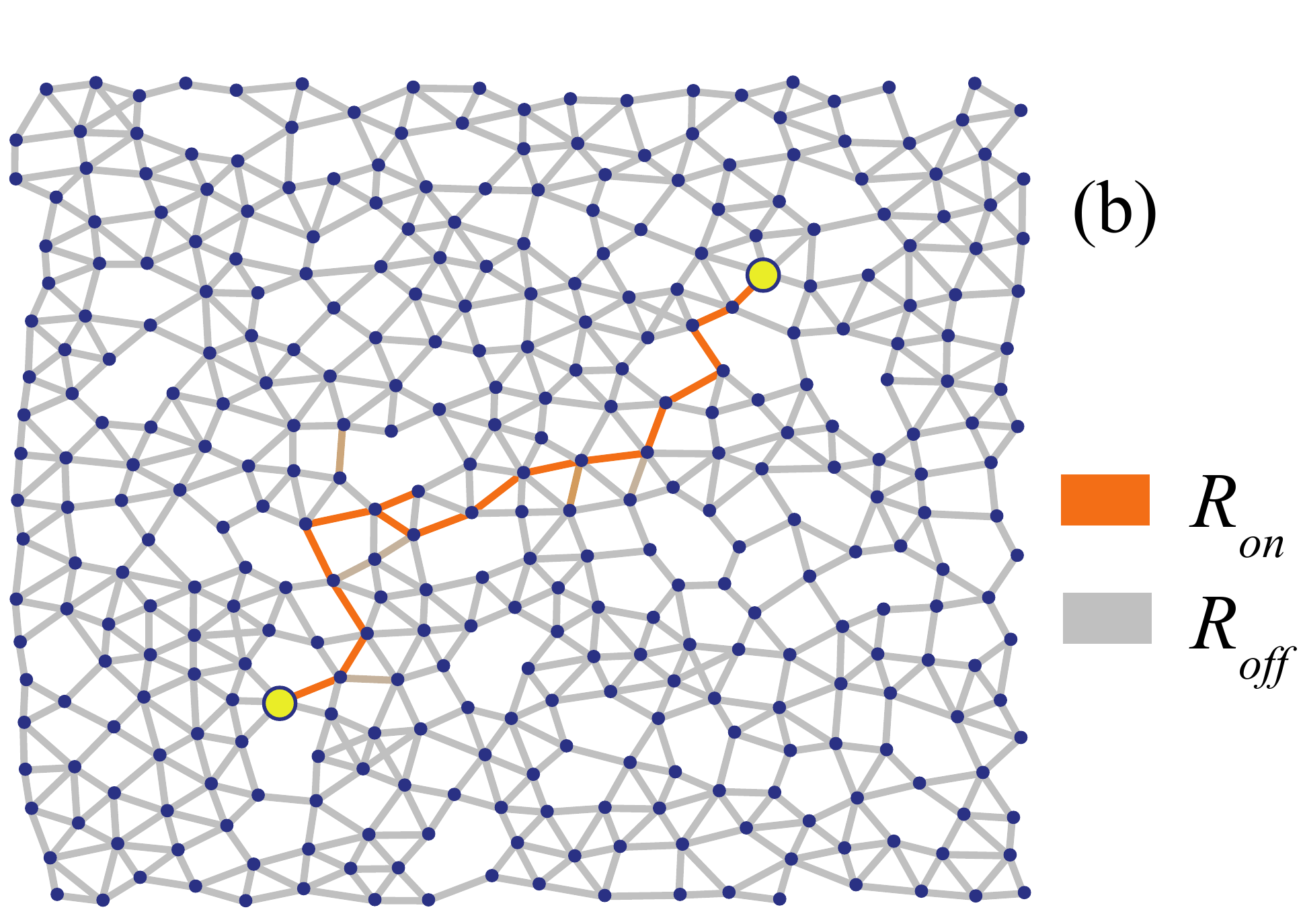}
\caption{\label{fig7} (Color online). Solution of the shortest-path problem by a random $20a\times 20a$ network for a horizontal (a) and diagonal (b) choice of the specified nodes (large yellow circles). These solutions are obtained for $R^M_{off}/R^M_{on}=100$.}
\end{center}
\end{figure}

Fig. \ref{fig7} shows solutions of the shortest-path problem found with random networks for two different choices of specified nodes. In both cases, the
solution is very close to the  shortest geometrical solution. However, in some cases,  small deviations from the shortest solution are possible such as a small 'bump' in Fig. \ref{fig7}(b) starting at the second node of the solution path bypassing an area of closely-spaced nodes (follow the solution path from the left bottom specified node). We associate such deviations with local inevitable fluctuations of network resistivity. As, initially, each edge in the network is of the same resistance, longer edges are more likely to be involved in the solution path as they offer less resistance to the current flow.

Clearly, a solution degeneracy or near-degeneracy is also possible with random networks. For example, Fig. \ref{fig7}(a) demonstrates a solution splitting in its central part.  In some cases, the solution
near-degeneracy (for example, 2 parallel segments of different length) can be removed by a post-processing
procedure in which the first problem solution (the set of edges in the $R_{on}$ state) is used as a new network for the shortest-path problem that is subsequently solved
following exactly the same algorithm as outlined in Sec. \ref{sec121}. Additionally, such post-processing procedure (that actually can be applied several times) removes
all possible 'dead ends' as well as segments disconnected from the solution path that occasionally are switched into $R_{on}$ state (an example of 'dead end' is the edge to above the left specified node in Fig. \ref{fig7}(a)).

\subsection{Traveling salesman problem} \label{sec32}

\begin{figure}[bt]
 \begin{center}
    \includegraphics[width=7.5cm]{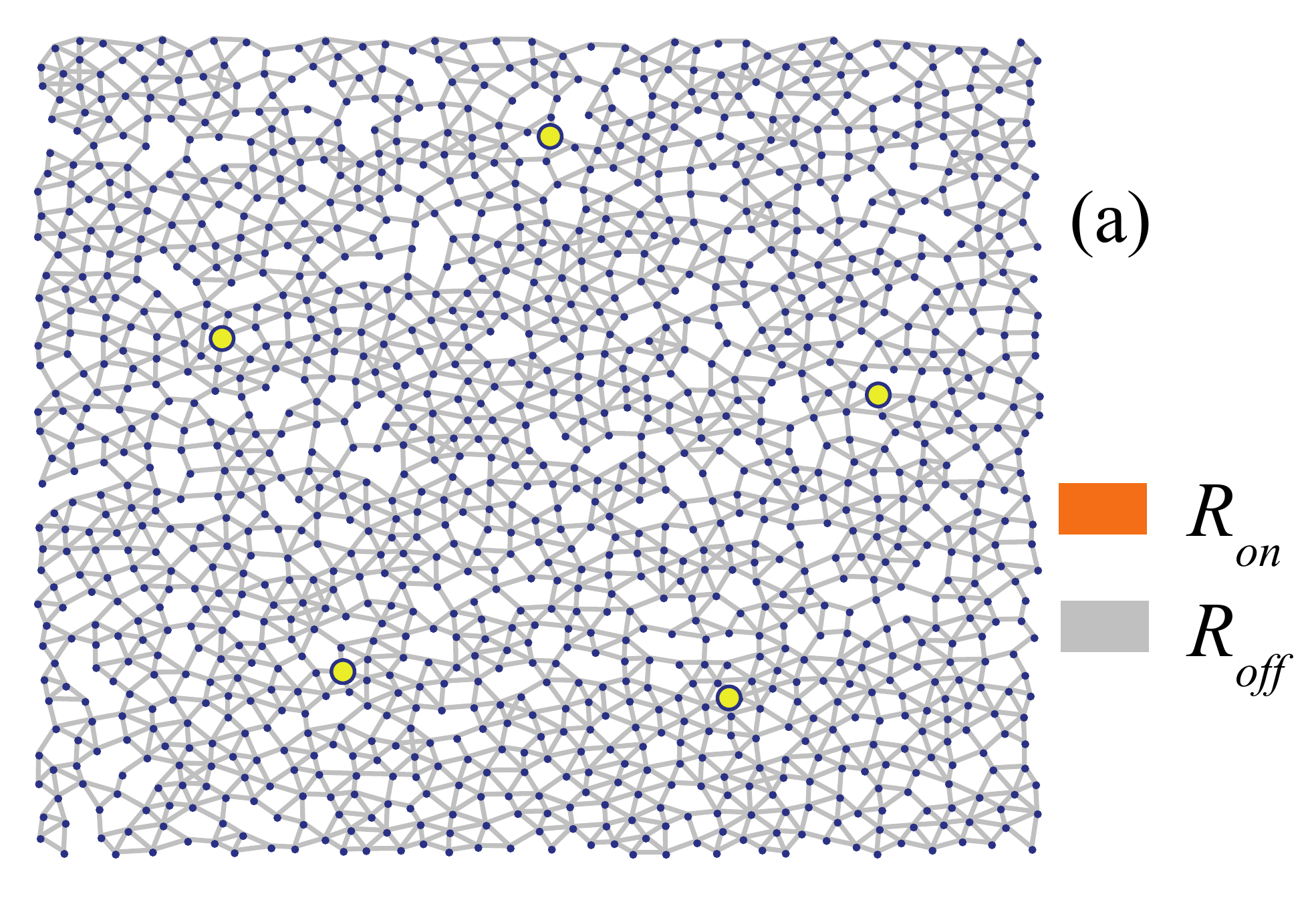}
    \includegraphics[width=7.5cm]{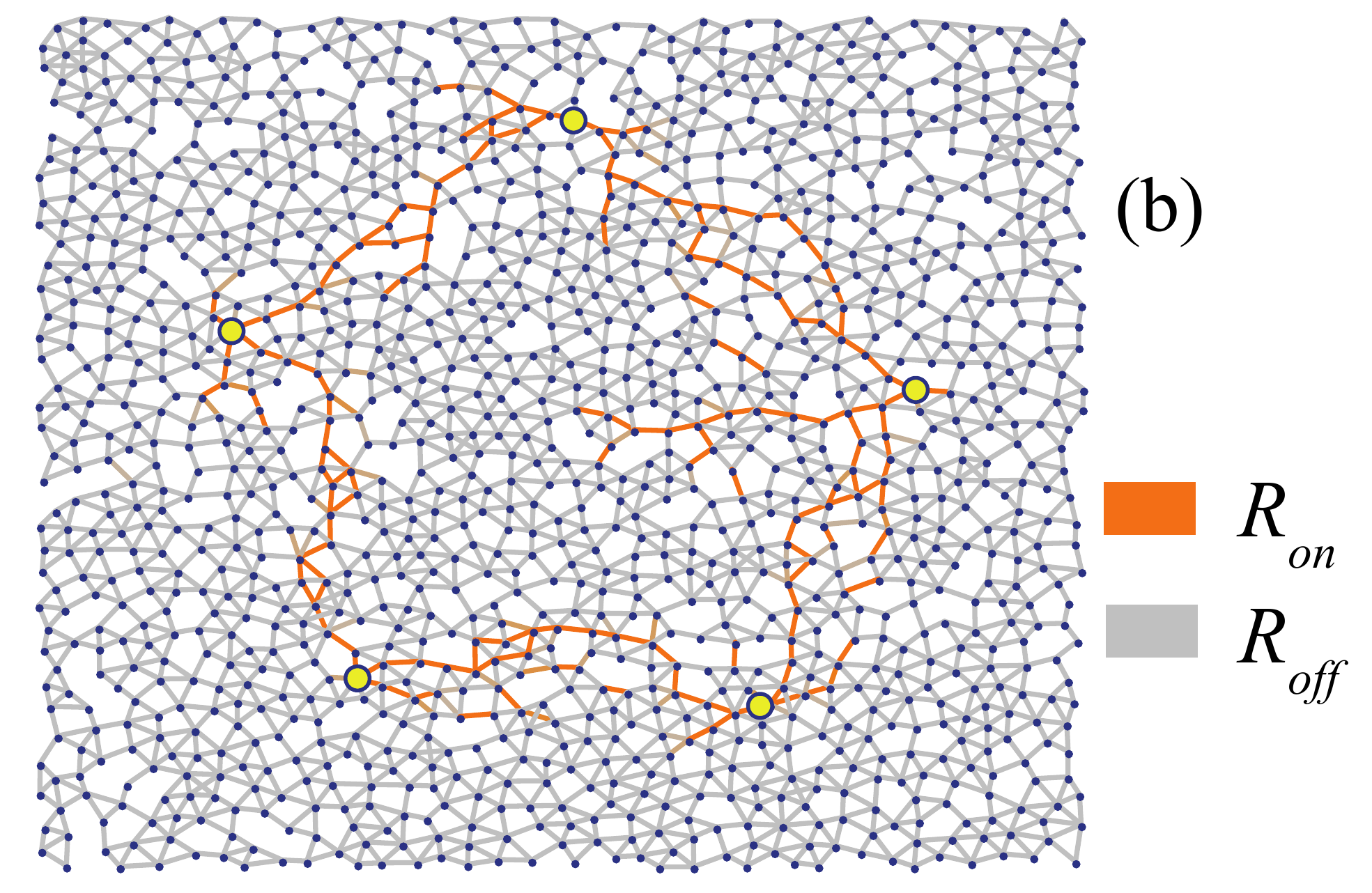}
    \includegraphics[width=7.5cm]{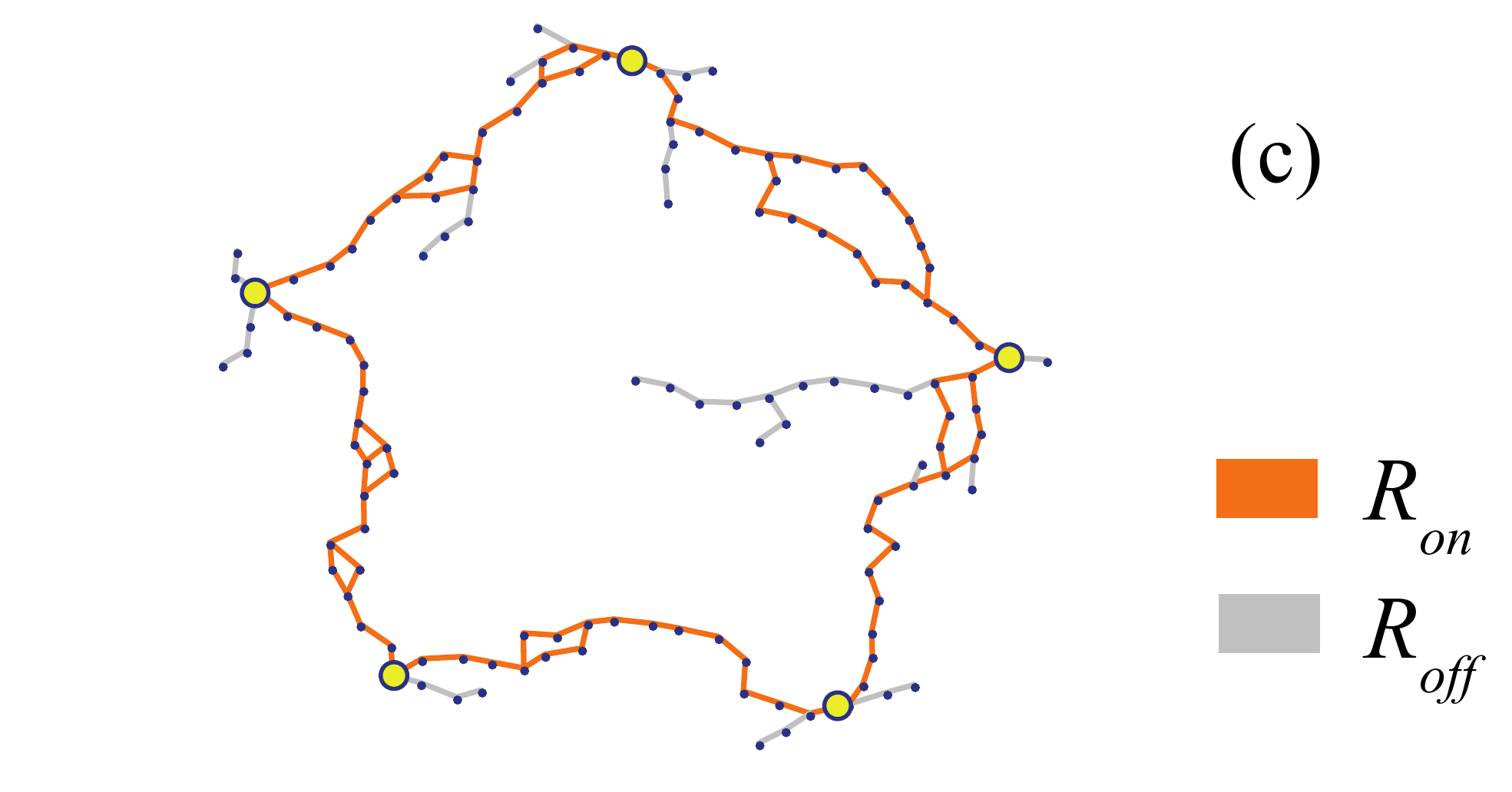}
\caption{\label{fig8} (Color online). Solving the traveling salesman problem using a random $40a\times 40a$ network: (a) initial state of the network with cities denoted by large yellow circles, (b) initial solution,
(c) improved solution (red lines) after post-processing where the `dead ends' (grey lines) have been
eliminated. This solution is obtained for a network with $R^M_{off}/R^M_{on}=100$. The final solution of the traveling salesman problem is represented by edges in $R_{on}$ state in (c).}
\end{center}
\end{figure}

Let us now consider the traveling salesman problem. For this purpose we use the algorithm described in Sec. \ref{sec122}, and we apply it twice. First of all, we apply this algorithm to the original network (Fig. \ref{fig8}(a))
and find the solution (that we call as the initial solution {\bf S1}) presented in Fig. \ref{fig8}(b). This solution, however, contains some disconnected edges in the $R_{on}$ states and 'dead ends' (paths with single connection). In order to improve on the initial solution, we use a post-processing procedure consisting in the following. We form a reduced network consisting of all edges of {\bf S1} that are in the $R_{on}$ state and nodes connected by these edges. Then, the traveling salesmen algorithm (Sec. \ref{sec122}) is applied to the reduced network. The total number of post-processing  operations depends on the network complexity. Networks presented in this paper normally require 1-2 operations of this type. The final solution is represented in Fig. \ref{fig8}(c) by edges in $R_{on}$ state (connected by red lines in the same figure). This is the correct solution of the traveling salesman problem for the given configuration of cities.

Currently, however, it is difficult to identify strict mathematical conditions when the traveling salesman problem can be correctly solved by our approach and when it can not be solved. But, even in the cases when the complete solution can not be found, it is quite possible that the suggested algorithm will provide partial solutions and thus tremendously reduce the number of possibilities to verify. Understanding these limitations will be subject of a future study.

\section{Conclusion} \label{sec4}

In conclusion, we have studied the possibility to solve shortest-path optimization problems with memristive networks.
Such approach is related to the concept of {\it memcomputing}: storing and processing of information on the same
physical platform~\cite{diventra13a}. There are six main criteria that need to be satisfied in order to realize such a paradigm
and it is clear that memristive networks considered in this paper satisfy all of them. Unlike other promising but more speculative proposals, like quantum computing, memcomputing
is already a practical reality, at least in regard to some applications, such as digital logic. It bypasses several
of the bottlenecks of present-day computing architectures and its constitutive units--memristors, memcapacitors, and
meminductors--are already widely available. Indeed, these elements emerge quite naturally with increasing  miniaturization of electronic devices. The computational possibilities offered by this paradigm are varied, and due to its tantalizing similarities both with some features of the brain as well as with the collective properties of colonies of living organisms, it promises to open new directions in neuromorphic architectures and biological studies.

In this work we have suggested and studied the implementation of the shortest-path and traveling salesmen problem algorithms utilizing memristive networks.
The main advantage of our approach is based on the analog parallel dynamics of many memristive devices. We have reported certain limitations of periodic networks
that in some cases provide degenerate solutions. Random and quasi-periodic networks without any (or a few) symmetries are more promising in this regard.
We have considered the network evolution that can be conveniently described by a network entropy. We anticipate that networks with memory have a wide range of applications
beyond the shortest-path problems. In particular, such networks can be used to simulate problem solving abilities of the human brain, the effect and progress of mental illness, etc.

\section{Acknowledgment}

This work has been partially supported by NSF grants No. DMR-0802830 and ECCS-1202383, and the Center for Magnetic Recording Research at UCSD.

\appendix

\section{Simulations}\label{cal_details}

The numerical results presented in this paper have been obtained for a network of current-controlled bipolar memristive devices with threshold. Each memristive device is described by
\begin{equation}
V_M=R\left(x \right)I_M, \label{Icontr4}
\end{equation}
and
\begin{numcases} {\frac{\textnormal{d}x}{\textnormal{d}t}=}
0 & for $|I_M|<I_t$  \label{Icontr5}
\\
\textnormal{sgn}\left( I_M\right)\gamma\left(\left|I_M\right|-I_t\right) & for $|I_M|\geq I_t$ \;\;\;\; \label{Icontr6}
\end{numcases}
where $V_M$ and $I_M=\dot{q}(t)$ denote the (time-dependent) voltage and current across the
device, respectively, $R(x)\equiv x$ is the memristance that changes between two limiting values $R^M_{on}$ and $R^M_{off}$,
$x$ is the internal state variable, $\gamma$ is a constant describing the rate of change of memristance when
the magnitude of the electric current $I_M$ exceeds the threshold current $I_t$; and $\textnormal{sgn}$ is the sign function.
We note that a current-controlled threshold-type memristive device model was used to describe switching in bipolar memristive devices \cite{Pickett09a}.
Moreover, many models of voltage-controlled memristive devices can be easily reformulated in the current-controlled form \cite{pershin11a}.
Eqs. (\ref{Icontr4})-(\ref{Icontr6}) was directly used in our studies of the shortest path problem. In the case of the traveling salesman problem, an absolute value of voltage was applied to each basic unit (edge) in order
to avoid back-and-forth switchings when random polarity voltage is applied. Such an approach can be easily implemented in electronics.

All numerical results were obtained for the following model parameters: $R^M_{off}=200$ Ohms, $R_{ij}(t=0)=R^M_{off}$, $\gamma=10^6$Ohms/(s$\cdot A$), $I_t=10$mA. Figs. \ref{fig2}, \ref{fig3}(a), \ref{fig5} and \ref{fig6} are obtained with $R^M_{on}=10$ Ohms and $V=6$V of applied voltage; Fig. \ref{fig3}(b) is found using $V=6,6.75,10,15.25$V for $R^M_{off}/R^M_{on}=20,10,4,1.25$ curves, respectively; Fig. \ref{fig4} is plotted with $R^M_{on}=160$ Ohms and $V=15.25$V. Note that $R^M_{on}$ and $R^M_{off}$ are
related to individual memristive devices, while $R_{on}$ and $R_{off}$ (used in figures) represent limiting values of memristance of the basic unit. While the "OFF" state of the basic unit is attained when both memristive devices are in their "OFF" states, the "ON" state of the basic unit corresponds to the "ON", "OFF" combination of single device states.
In our simulations, at each time step, the potential at all grid points is found as a solution of Kirchhoff's current law equations obtained using a sparse matrix technique. The corresponding change in the memristive states was computed using Eq. (\ref{Icontr6}). The width of the voltage pulse is selected sufficiently long to reach the steady state in each calculation.

\bibliography{memcapacitor}
\end{document}